# Determination of bound state energies for a one-dimensional potential field

D.M.Sedrakian [1], A.Zh.Khachatrian [2]

[1] Yerevan state university 375025, Yerevan, Armenia, Alex Manukyan 1,

dsedrak@www.physdep.r.am

[2] The state engineering university of Armenia, 375046 Yerevan, Armenia, Teryan 109,

akhachat@www.physdep.r.am

A method for determination of bound state energies for an asymmetric quantum well with an arbitrary shape of the bottom is suggested. It is shown that how the equation determining the energy levels can be easily derived if one knows the electron transmission and reflection amplitudes corresponding to the part of potential inside the well. The results are applied to three difference test problems.

## Introduction

It is well known that the description of elementary excitations and their energy spectra in one - dimensional non-regular systems is of physical interest and plays an important role in different problems of the modern applied physics. Recently, experiments on nano-size layered structures have lead to a greater interest in the study of a quantum particle's motion, which moves into the limited volume contacting inside an arbitrarily changing from point to point potential [1-3].

Here we consider a problem of determination of energy levels of bound states in an arbitrary one-dimensional potential well. Let us consider the electron's motion in the field of a potential having the following form:

$$\left(-\frac{\hbar^2}{2m}\frac{d^2}{dx^2}+V(x)\right)\psi(x)=E\psi(x), \qquad (1)$$

$$V(x)=\begin{cases} V_1=const, & x\leq x_1, \\ V(x), & x_1<x<x_2, \\ V_2=const, & x\geq x_2 \end{cases} \qquad (2)$$

where, in general, the magnitudes of $V_1$, $V_2$ are different and $V(x)$ is an arbitrary function of $x$. If one takes the zero of energy to be equal to the minimum of the potential $V(x)$, then the energy spectrum will be continuous provided that $V_1=V_2=0$.

This problem can be found in many quantum mechanic text book. A similar problem to in optics is the problem of eigen frequency determination for a plane polalized electromagnetic field in an asymmetric waveguide with a non-homogeneous layered structure [4]. In this paper, we present a new general approach for solving such problems. It should be mentioned at the outside that the main difficulty of the problem is the determination of energy levels. If the energy spectrum is known, the construction of the wave functions can be done by any standard method, e.g. the transfer matrix approach [5].

Because of the state localization on surfaces or in the bulk of a complex structure, at present finite periodic structures are more studied, when in two semi-infinite media an ideal periodic structure exists [6-9]. In this work, we introduce an approach for solving the problem of bound state determination for the one-dimensional potential well with an arbitrary shape of the bottom. We show that if the scattering problem for the part of the potential located in the well has been solved, then the equation determining the bound state energies can be straightforwardly found.

## 2. Spectrum equation

The bound states are solutions of Eq. (1) damping to zero in the intervals $(-\infty, x_1)$ and $(x_2, \infty)$ ;

$$\psi(x) = \begin{cases} D\exp\{\chi_1 x\}, x < x_1, \\ C\exp\{-\chi_2 x\}, x > x_2, \end{cases} \quad (3)$$

where $V_1 - E = \hbar^2 \chi_1^2 / 2m$ и $V_2 - E = \hbar^2 \chi_2^2 / 2m$.

The conditions of continuity of the wave function and its derivation at boundary points $x_1$ and $x_2$ determine the energy spectrum of the bound states. Therefore one can write:

$$\begin{cases} D\exp\{\chi_1 x_1\} = \psi(x_1), \quad \chi_1 D\exp\{\chi_1 x_1\} = d\psi(x_1)/dx, \\ C\exp\{-\chi_2 x_2\} = \psi(x_2), \quad -\chi_2 C\exp\{-\chi_2 x_2\} = d\psi(x_2)/dx, \end{cases} \quad (4)$$

where $\psi(x)$ is the general solution of wave equation (1) with potential $V(x)$.

Let us suppose, that near the boundary points, there are infinitely small regions, where the magnitude of the potential equals to zero. In these regions the wave functions can be written in the following form:

$$\psi(x) = \begin{cases} A_1 \exp\{ikx\} + B_1 \exp\{-ikx\}, & x_1 < x < x_1 + 0, \\ A_2 \exp\{ikx\} + B_2 \exp\{-ikx\}, & x_2 - 0 < x < x_2, \end{cases} \quad (5)$$

where $E = \hbar^2 k^2 / 2m$.

According to the transfer matrix approach, the following linear relationship between the coefficients of the solution exists [5]:

$$\begin{pmatrix} A_2 \\ B_2 \end{pmatrix} = \begin{pmatrix} \alpha & \beta \\ \beta^* & \alpha^* \end{pmatrix} \begin{pmatrix} A_1 \\ B_1 \end{pmatrix}, \quad (6)$$

where $\alpha = 1/t^*$, $\beta = -r^*/t^*$ with $r$ and $t$ being the electron reflection and transmission amplitudes for the part of the potential between the points $x_1$ and $x_2$, when the potential is zero at the both boundaries ($V_1$, $V_2 = 0$).

In the following, we take $x_1 = 0$ and $x_2 = d$. By using expression (6), condition (4) can be rewritten as:

$$D = A_1 + B_1, \quad (7)$$

$$D = \frac{ik}{\chi_1}(A_1 - B_1), \quad (8)$$

$$C \exp\{-\chi_2 d\} = (\alpha \exp\{ikd\} + \beta^* \exp\{-ikd\})A_1 + (\beta \exp\{ikd\} + \alpha^* \exp\{-ikd\})B_1, \quad (9)$$

$$C \exp\{-\chi_2 d\} = \frac{ik}{\chi_2}\left[(\beta^* \exp\{-ikd\} - \alpha \exp\{ikd\})A_1 + (\alpha^* \exp\{-ikd\} - \beta \exp\{ikd\})B_1\right]. \quad (10)$$

Let us consider Eq. (7)-(10) as a linear set of homogenous equations with unknown quantities $A_1, B_1, C, D$. The requirement of the set determinant to be equal to zero gives

$$\begin{vmatrix} \chi_1 - ik & \chi_1 + ik \\ (\chi_2 + ik)\alpha e^{ikd} + (\chi_2 - ik)\beta^* e^{-ikd} & (\chi_2 + ik)\beta e^{ikd} + (\chi_2 - ik)\alpha^* e^{-ikd} \end{vmatrix} = 0. \quad (11)$$

Eq.(11) can be written in the following form:

$$tgkd = \frac{k(\chi_1 + \chi_2)\operatorname{Re}\alpha + (\chi_1\chi_2 - k^2)\operatorname{Im}\alpha - (\chi_1\chi_2 + k^2)\operatorname{Im}\beta - k(\chi_1 - \chi_2)\operatorname{Re}\beta}{k(\chi_1 + \chi_2)\operatorname{Im}\alpha - (\chi_1\chi_2 - k^2)\operatorname{Re}\alpha + (\chi_1\chi_2 + k^2)\operatorname{Re}\beta - k(\chi_1 - \chi_2)\operatorname{Im}\beta}. \quad (12)$$

Eq. (12) is an analytical condition for determination of the electron bound states in the asymmetric quantum well with an arbitrary shape of the bottom. It follows from Eq. (12), that the problem of determination of the bound state energies is reduced to the problem of determination of the scattering amplitudes $r$ and $t$. The latter was treated by using difference methods and, in general, it can be solved only numerically (see e.g. [10, 11] and references therein). Note that in Eq. (12), quantities $\alpha$ and $\beta$ are the functions of energy as well, which, as it was mentioned above, can be determined from the electron scattering problem for the part of potential Eq.(1), with in the well.

From Eq.(12), when $\chi_1 = \chi_2 = \chi$, we get the following equation determining the bound state energies for the symmetric well with an arbitrary bottom:

$$tgkd = \frac{2k\chi \operatorname{Re}\alpha + (\chi^2 - k^2)\operatorname{Im}\alpha - (\chi^2 + k^2)\operatorname{Im}\beta}{2k\chi \operatorname{Im}\alpha - (\chi^2 - k^2)\operatorname{Re}\alpha + (\chi^2 + k^2)\operatorname{Re}\beta}. \qquad (13)$$

It is interesting to consider how Eq.(13) reduces to the well known equation, which determines the energy spectrum for a well with a plane bottom. Inserting into Eq.(13) $\alpha = 1$, $\beta = 0$ we get

$$tgkd = \frac{2k\chi}{k^2 - \chi^2}. \qquad (14)$$

Substituting $tgkd = 2tg(kd/2)/(1 - tg^2(kd/2))$ into Eq.(13) and solving then quadratic equation with respect to $\tan(kd/2)$, we get [12]

$$\tan\frac{kd}{2} = \frac{\chi}{k} \text{ и } ctg\frac{kd}{2} = -\frac{\chi}{k}. \qquad (15)$$

The first and second equations determine the energy levels for the symmetric and asymmetric electron states for a simple well of width $d$.

When $\chi \to \infty$ Eq.(13) to infinity, the spectrum equation for an infinite quantum well with an arbitrary shape of the bottom is

$$tgkd = \frac{\operatorname{Im}\alpha - \operatorname{Im}\beta}{\operatorname{Re}\beta - \operatorname{Re}\alpha}. \qquad (16)$$

For the plane bottom well ($\alpha = 1$, $\beta = 0$) Eq.(16) takes on the well known form

$$\sin kd = 0 \text{ и } k_n = \frac{\pi}{d}n, \qquad (17)$$

where $n = 1, 2, 3 \cdots$.

### 3. A single $\delta$-like potential in a symmetric well

It is interesting to apply the above obtained results to the problem of energy spectrum determination, when the well contains a single $\delta$ - like barrier. Denoting by $d_1$ ($d_1 < d$) the distance between the $\delta$ - barrier and the left wall of the well, and using the well known expressions for transmission and reflection amplitudes of $\delta$ - barrier we find:

$\operatorname{Re}\alpha = 1$, $\operatorname{Im}\alpha = -V/2k$, $\operatorname{Re}\beta = -V\sin\{2kd_1\}/2k$, $\operatorname{Im}\beta = -V\cos\{2kd_1\}/2k$, (18)

where $V$ is the power of $\delta$-potential. By inserting Eq.(18) in Eq.(13), the latter can be written in the form of

$$\sin\{kd_1\}\sin\{kd_2\}\left\{\frac{B+Au}{C}-\frac{B-Au}{C}ctg\{kd_2\}ctg\{kd_1\}+ctg\{kd_2\}+ctg\{kd_1\}\right\}=0, \quad (19)$$

where the following notations are used

$$A=1+\left(\frac{k}{\chi}\right)^2, B=\left[1-\left(\frac{k}{\chi}\right)^2\right]u-\frac{2k}{\chi}, C=\left[1-\left(\frac{k}{\chi}\right)^2\right]-\frac{V}{\chi}, u=\frac{V}{2k}, d_2=d-d_1. \quad (20)$$

This equation determines the energy spectrum for a symmetric well inside of which $\delta$-potential is located at an arbitrary point.

As follows from (19), when $B-Au\neq 0$ the spectrum is determined by one equation by:

$$\frac{B+Au}{C}-\frac{B-Au}{C}ctg\{kd_2\}ctg\{kd_1\}+ctg\{kd_2\}+ctg\{kd_1\}=0. \quad (21)$$

When $B-Au=0$ (It particularly can take place for the case of infinite depth well), the spectrum is determined by one of the following equations:

$$\sin\{kd_1\}\sin\{kd_2\}=0 \quad \text{or} \quad -\frac{2uA}{C}=ctg\{kd_2\}+ctg\{kd_1\}. \quad (22)$$

The first equation can have a solution when the ratio of $d_2$ and $d_1$ is a rational number ($d_1/d_2=N/M$, where $N,M$ are natural numbers). When this ratio is not a rational number, then the spectrum is determined by the second equation of Eq. (22) only.

For the case, when the $\delta$- like barrier is located at the center of the infinite depth well ($d_1=d_2=d/2$ and $2uA/c=V/k$), the equations (22) take the known form [13]

$$\sin\{kd/2\}=0, \quad ctg\{kd/2\}=-V/2k. \quad (23)$$

Note that as follows from the first equation of (23) the $\delta$- like barrier does not change the values of energy levels for even states.

It is interesting to mention that the result Eq.(12) can be found by mains of solving the scattering problem for the potential Eq.(2). As it was shown in the Ref. [14], the transmission amplitude of an electron incident on the potential Eq.(2) from the left can be presented in the form of :

$$\frac{1}{t_{I,II}} = \frac{\exp\{ik_2 d\}}{4k_1 k}\left[\frac{(k_2-k)(k-k_1)}{t^*}\exp\{ik_0 d\} + \frac{(k_2+k)(k+k_1)}{t}\exp\{-ik_0 d\} + \right.$$
$$\left. +\frac{(k+k_2)(k_1-k)r}{t}\exp\{-ik_0 d\} + \frac{(k-k_2)(k+k_1)r^*}{t^*}\exp\{ik_0 d\}\right], \qquad (24)$$

where $E - V_1 = \hbar^2 k_1^2/2m$, $E - V_2 = \hbar^2 k_2^2/2m$. Here $r$ and $t$ are the transmission and reflection amplitudes when the value of the potential at the left and right boundaries of the layer equals to zero $V_1 = V_2 = 0$. As it is known, the poles of the transmission amplitude determine the energy spectrum of bound states. Indeed, taking in Eq.(24) $k_1 = i\chi_1$, $k_2 = i\chi_2$ and considering the equation $1/t_{I,II} = 0$, it is easy to see that the last coincides with the equation (12).

### 4. A rectangular potential in a symmetric well

Further we apply the result Eq.(12), when a rectangular barrier is located inside of the rectangular well. As it was mentioned above, the equation determining the energy levels is a complicated transcendental equation. Because of this, the problem of determination of the spectrum equation's roots as a function of barrier parameters (the magnitude of the potential, the width, and its position) can be solved only numerically. For this problem, the quantities $\alpha$, $\beta$ have the following form:

$$\alpha = \exp\{-ik\Delta\}\left\{\cos(q\Delta) + i\frac{k^2+q^2}{2kq}\cos(q\Delta)\right\}, \qquad (25)$$

$$\beta = i\frac{k^2-q^2}{2kq}\sin(q\Delta)\exp\{-i2k(a+\Delta/2)\}. \qquad (26)$$

where $\Delta$ is the barrier width, $a$ is the distance between the barrier left wall and the left wall of the well (note, that $a+\Delta/2$ is the coordinate of the barrier middle point), $q^2 = E - U$ and $U$ is the magnitude of the barrier potential.

In the case of a symmetric quantum well for the first three energy levels, the results of the corresponding calculations are given on Figures 1 and 2. The following quantities are chosen as the dimensionless parameters of problem

$$E_1 = \varepsilon_1 \Delta^2, E_2 = \varepsilon_2 \Delta^2, E_3 = \varepsilon_3 \Delta^2, V_1 \Delta^2 = V_2 \Delta^2 = v, u = V\Delta^2, x = \Delta/d, \; y = a/\Delta, \quad (26)$$

where $2m\varepsilon_1/\hbar^2, 2m\varepsilon_2/\hbar^2, 2m\varepsilon_3/\hbar^2$ correspond to the first three energy levels.

In Figure 1, the graphs of $E_1, E_2, E_3$ and $u$ for different values of $x$ are plotted, at $v = 5$ and $y = 0$ (the barrier attached to the well wall). The value of dimensionless parameter $x$ was chosen in such a way that the three levels to be in the well only. Note that at $x = 0.2$ in the empty well the number of bound states equals to four. According to the curves given in Figure 2, the increase of $u$ leads to the increase of $E_1, E_2, E_3$. This means that for a fixed value of the barrier width the increasing of the barrier height throws the levels out of the well. It is seen from Figure 1 that at $x = 0.3$ the third energy level transforms into the continuum spectrum for $u > 3.1$.

Due to the problem of the second harmonic generation in the complex quantum wells [15-17], it is interesting to discuss the question of the energy levels equidistant with the well ($E_3 - E_2 = E_2 - E_1$). This problem for the infinite quantum well was considered in Ref. [15], where, in particular, it was shown that for a given value of equidistance the ensemble of potentials protecting this equidistance exists. As it can be seen from Figure 1 and as the calculations show, the equidistance between the levels when x equals 0.2 and 0.25 is impossible. At $x = 0.3$, the levels are equidistant for $u = 1.845$ and it is seen from the curves that the energy value of the main first level is less than the value of the barrier potential, and the second and third levels are located above the barrier. It is important to note that for an infinite deep well the equidistant between the three lowest levels is possible since $x > 0.34$. This means, that a more realistic case for the finitely deep well the equidistance of levels can be reached with the help of barrier having moreover width, then for the case of infinite well.

In Figure 2, the dependencies of $E_1, E_2, E_3$ on $y$ for different values of $u$ and for fixed values of $x = 0.3$, $v = 5$ (the empty well with these values of the parameters has the three bound states) are given. Note that for a fixed value of the barrier the quantity $y$ describes the location of the barrier in the well. So for $y = 0$, the barrier attached to the well's left wall and for $y = 1/x - 1$ (for chosen $x$ we get $y = 2.333$) it is attached to the well's right wall. As it is seen from Figure 2 the dependencies of level locations are symmetric with respect to the point $x = (1/x - 1)/2$, which coincides with the well's middle point. The latter is connected with the empty well symmetry.

It is interesting to note, that the first level depending on the barrier location has one minimum, the second level has one minimum and two maxima and the third one has two maxima and one minimum. Such dependencies of levels are easy to understand if one consider the location of the extrema of the wave functions for the empty well.

### 5. A periodic layered structure in a infinite well

In this section, we consider the energy spectrum for an infinitely deep well containing an ideal periodic structure with $N$ rectangular potential barriers inside. According to Eq. (16), the equation determining the energy levels can be written in the following form:

$$\sin\{kd\}(\operatorname{Re}\{R_N/T_N\} + \operatorname{Re}\{1/T_N\}) - \cos\{kd\}(\operatorname{Im}\{R_N/T_N\} + \operatorname{Im}\{1/T_N\}) = 0, \qquad (27)$$

where the transmission and reflection amplitudes $T_N$, $R_N$ of the ideal structure are given by [3] (see for example [18] as well)

$$\frac{1}{T_N} = \exp\{ikNa\}\left\{\cos N\beta + i\operatorname{Im}(t_1^{-1}\exp\{-ika\})\frac{\sin N\beta}{\sin \beta}\right\}, \qquad (28)$$

$$\frac{R_n}{T_n} = \exp\{ik(N-1)a\}\frac{r_1}{t_1}\frac{\sin N\beta}{\sin \beta}, \qquad (29)$$

where $\cos\beta = \operatorname{Re}(\exp\{-ika\}/t_1)$ ($t_1$, $r_1$ are the transmission and refection amplitudes for the first potential of the ideal structure, $a$ is the structure period), which for the ideal structure composed of rectangular potentials has the form of

$$\cos\beta = \cos k(a-l)\cos ql - \frac{q^2+k^2}{2kq}\sin k(a-l)\sin ql, \qquad (30)$$

$$\frac{1}{t_1} = \exp\{ikd\}\left\{\cos qd - i\frac{k^2+q^2}{2qk}\sin qd\right\}, \qquad (31)$$

$$\frac{r_1}{t_1} = i\exp\{ikx_1\}\frac{k^2-q^2}{2qk}\sin qd, \qquad (32)$$

where $l$ is the width of the single rectangular barrier, $q^2 = E - U$ and $U$ is the magnitude of the barrier potential, $x_1$ is the coordinate of the middle point of the first potential.

In Figure 3, dependence of the energy levels on the value of the rectangular barriers is shown, when $a/d = 10$ and inside the infinite well $n = 3, 4, 5, 6, 7, 8$ rectangular barriers are located. As seen from Figure 3, the increase of the potential value of the rectangular barrier leads to the band structure of the spectrum. The levels merge into the bands. Note that each of the bands at a fixed number of $N$ contains $N+1$ band states.

**Conclusion**

In this paper we introduce a new approach to the bound state problem for a one-dimensional asymmetric well with an arbitrary shape of the bottom. As we have shown, the equation determining the energy levels can be easily rewritten if the transmission and reflection amplitudes for the part of the potential located inside the well are known. The latter problem can be generally solved only numerically. For example, in the case of a layered structure, this problem is generally reduced to the solution of some linear finite-difference equations [10]. In the case of a continuous potential, this problem can be formulated as the Cauchy problem for the wave equation [11].

It is also shown that the problem of determination of bound state energies can be considered with the help of the scattering problem for an electron incident on the potential well. The poles of the transmission amplitude also determine the energy levels. The suggested method is applied for consideration of three specific cases. First of them is the symmetric quantum well containing a simple $\delta$-like potential. We have shown that when the $\delta$- potential is located at the knot of the wave function it does not change the energy level. The second case is a rectangular potential barrier located in the symmetric well. As we have shown, the difference between the lowest energy levels by the special choice of the potential parameter can be made equidistant. And in the last third case, we consider a finite quantum well containing a periodic structure of rectangular barriers. The dependence of the energy level positions on rectangular potential value is plotted. The increase of the potential leads to merging of the energy levels into the energy bands.

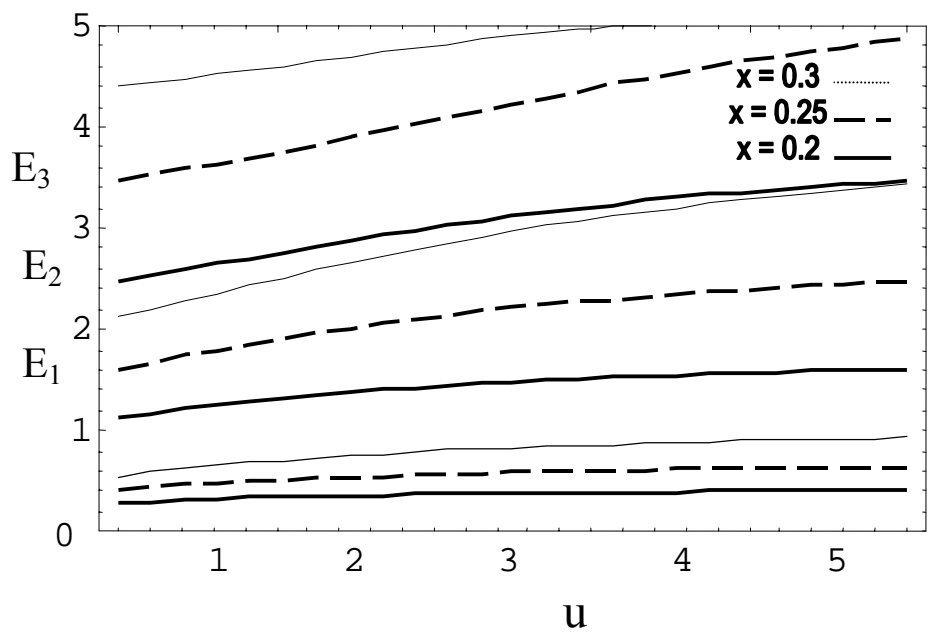

Figure 1. The dependencies of $E_1, E_2, E_3$ on $u$ for the different values of $x$ and at the fixed $v = 5$ and $y = 0$.

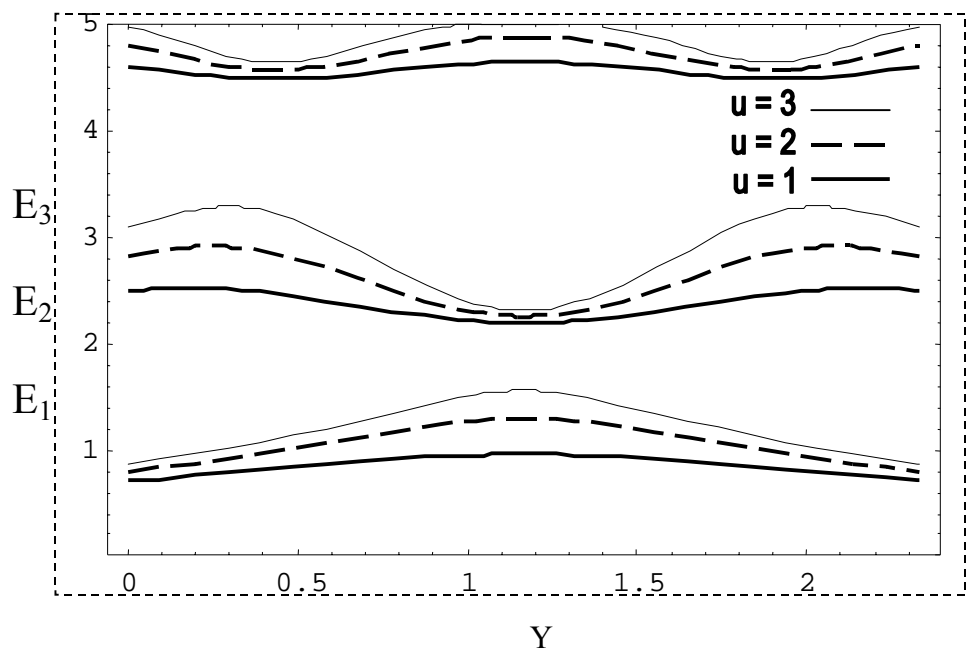

Figure 2. The dependencies of $E_1, E_2, E_3$ on $y$ for different values of $u$ and at fixed $x = 0.3$, $v = 5$.

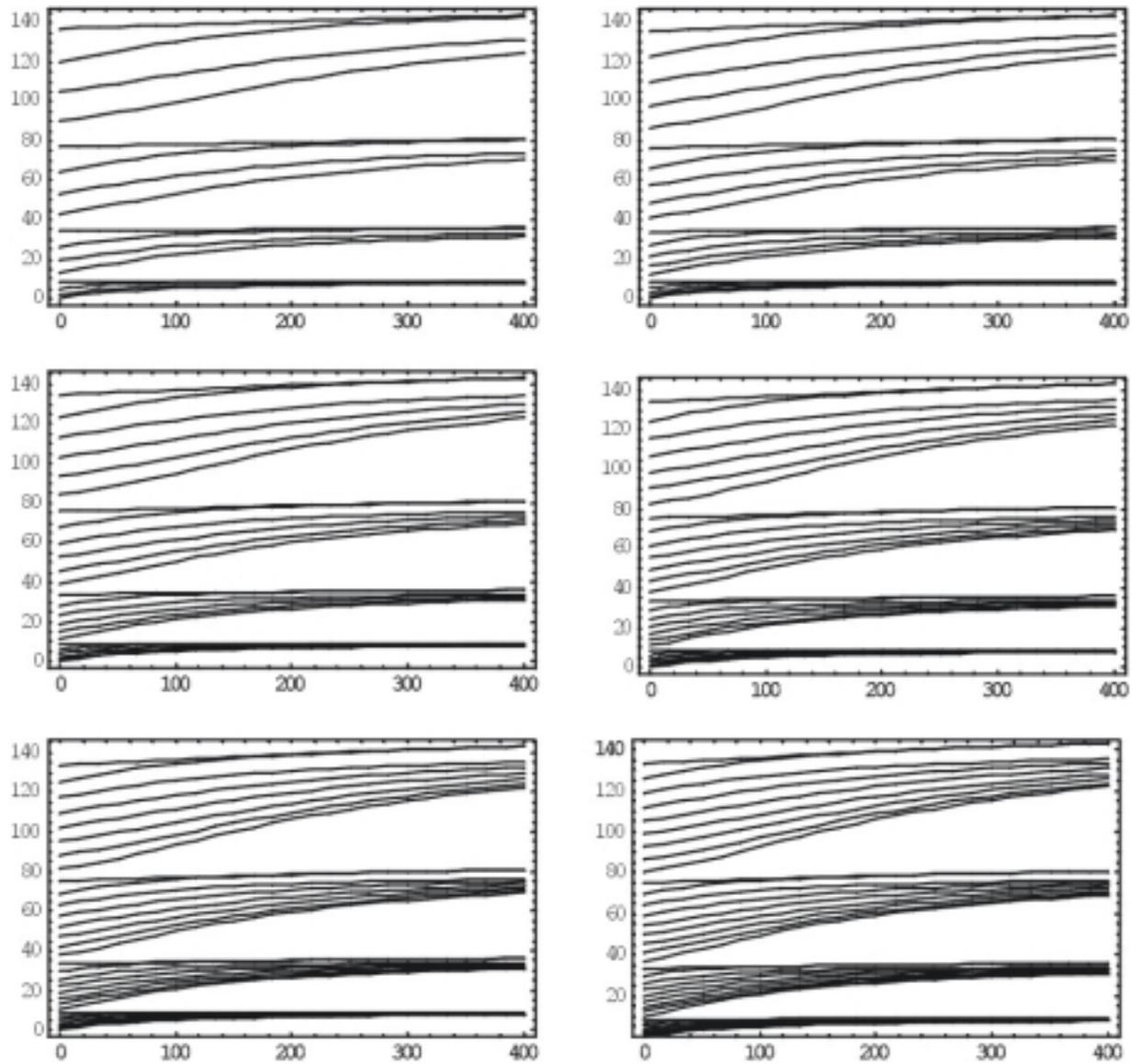

Figure 3. Dependence of the energy levels on the value of rectangular barriers. $a/d = 10$ and inside the infinite well $n = 3, 4, 5, 6, 7, 8$ rectangular barrieres are located.